\documentclass[preprint,3p]{elsarticle}
\usepackage{amsmath, amssymb, amsthm}
\usepackage{mathtools}
\usepackage{mathrsfs}
\usepackage{bm}
\usepackage{slashed}   %feynman slashed symbols
\usepackage{graphicx}
\usepackage{multirow}
\usepackage{tikz}
\usepackage[caption=false]{subfig}
\usepackage{relsize}	%rescalable math
\usepackage{array}
\usepackage{float}
\usepackage{color}
\usepackage{xcolor}
\usepackage{soul}
\biboptions{numbers,square,comma,sort&compress}
\usepackage{siunitx}
\usepackage{hyperref}
\hypersetup{
}
\def\be{\begin{eqnarray}}
\def\ee{\end{eqnarray}}
\mathchardef\-="2D

\usepackage{hyperref}

\hypersetup{
colorlinks=true,linkcolor=red,anchorcolor=blue,citecolor=blue, filecolor=blue,urlcolor=red,bookmarksnumbered=true, pdfview=FitB
}

\begin{document}
\begin{frontmatter}
\title{Nucleon electroweak form factors using spin-improved holographic light-front wavefunctions}

\author[1]{Mohammad Ahmady}

\author[2]{Dipankar Chakrabarti}

\author[3,4]{Chandan Mondal\corref{cor1}}
\ead{mondal@impcas.ac.cn}

\author[5]{Ruben Sandapen}

\address[1]{Department of Physics, Mount Allison University,  Sackville, New Brunswick, Canada, E4L 1E6}

\address[2]{Department of Physics,  Indian Institute of Technology Kanpur, Kanpur 208016, India}

\address[3]{Institute of Modern Physics, Chinese Academy of Sciences, Lanzhou 730000, China}
%\address[2]{CAS Key Laboratory of High Precision Nuclear Spectroscopy, Institute of Modern Physics, Chinese Academy of Sciences, Lanzhou 730000, China}
\address[4]{School of Nuclear Science and Technology, University of Chinese Academy of Sciences, Beijing 100049, China}

\address[5]{Department of Physics, Acadia University,  Wolfville, Nova-Scotia, Canada, B4P 2R6}

\cortext[cor1]{Corresponding author}

\begin{abstract}
We construct spin-improved holographic light front wavefunctions for the nucleons (viewed as quark-diquark systems) and use them to successfully predict their electromagnetic Sachs form factors, their electromagnetic charge radii, as well as the axial form factor, charge and radius of the proton. The confinement scale is the universal mass scale of light-front holography, previously extracted from spectroscopic data for light hadrons. With the Dirac and Pauli form factors normalized using the quark counting rules and the measured anomalous magnetic moments respectively, the masses of the quark and diquark are the only remaining adjustable parameters. We fix them using the data set for the proton's Dirac-to-Pauli form factor ratio, and then predict all other data without any further adjustments of parameters. 
Agreement with data at low momentum-transfer is excellent. Our findings support the idea that light (pseudoscalar and vector) mesons and the nucleons share a nonperturbative universal  holographic light-front wavefunction which is modified differently by their spin structures.
\end{abstract}

\begin{keyword}
Light-front holography, Quark-diquark model, Electroweak form factors, Nucleon
\end{keyword}

\end{frontmatter}

%=========================================
\section{Introduction}
\label{Intro}
%%%%%%%%%%%%%%%%

Previous papers \cite{Forshaw:2012im,Ahmady:2016ujw,Ahmady:2016ufq,Ahmady:2019yvo,Kaur:2020emh}, and especially Ref. \cite{Ahmady:2020mht}, support the idea that light pseudoscalar and vector mesons share a universal holographic light-front wavefunction modified by their different spin structures.  By the universal holographic wavefunction, we mean the $N$-quark valence wavefunction \cite{Brodsky:2013npa,Brodsky:2006uqa}
\begin{equation}
	\Psi(x,\zeta)= X(x) e^{iL\varphi} \frac{\phi(\zeta)}{\sqrt{2\pi \zeta}} \,,
\label{holographic-wfn-impact}
\end{equation}
where $\zeta$ is given by
\begin{equation}
	\mathbf{\zeta}=\sqrt{\frac{x}{(1-x)}}\left|\sum^{N-1}_{j=1} x_j \mathbf{b}_{\perp,j}\right| \,,
\end{equation}
where $N=2$ for mesons and $N=3$ for baryons. The variable $x$ is the light-front momentum fraction of one quark (labelled active) while $x_j$ are the momentum fractions associated with the $N-1$ remaining quarks (labelled spectator cluster). The variables $\mathbf{b}_{\perp,j}$ are the transverse positions of the spectator quarks relative to the active one.  The quantum number $L$ is the relative orbital angular momentum between the active quark and the spectator cluster. In a quark-antiquark meson, when the quark is active (spectator), the antiquark is spectator (active). Therefore, $\zeta=\sqrt{x \bar{x}} b_\perp$,  where $\bar{x} \equiv (1-x)$ and $b_\perp$ is the transverse separation between the quark and the antiquark. Eq.~(\ref{holographic-wfn-impact}) also describes a baryon in which the spectator quarks constitute a diquark cluster. In this case, when any one quark is active, it carries momentum fraction $x$, and the remaining two form a spectator diquark cluster which carries momentum fraction $\bar{x}$. 

The factorized form of Eq.~(\ref{holographic-wfn-impact}) is realized in the conformal limit of light-front QCD, i.e. when quantum loops and quark masses are neglected. In this limit, the transverse mode, $\phi(\zeta)$, satisfies the light-front holographic Schr\"odinger Equation \cite{Brodsky:2014yha}
\begin{equation}
	\left(-\frac{\mathrm{d}^2}{\mathrm{d}\zeta^2}-\frac{4L^2-1}{4\zeta^2} + U(\zeta, J) \right)\phi_J(\zeta)=M^2 \phi_J(\zeta) \,,
\label{hSE-mesons}
\end{equation} 
where $J$ is the meson's spin. Note that Eqs.~(\ref{holographic-wfn-impact}) and (\ref{hSE-mesons}) do not carry any information on the spins of the quarks. In light-front holographic QCD, this is a consequence of the underlying assumption that the quark helicity wavefunction decouples from the confinement dynamics. Eq.~(\ref{hSE-mesons}) is said to be holographic since it maps onto the equation of motion for bosonic modes, $\Phi_J(x,z)=\Phi_J(z)e^{iP \cdot x}$, propagating in a $\mathrm{AdS}_5$ spacetime modified by a dilaton field, $\varphi(z)$, according to the dictionary: 
 \be
 	&&\zeta = z\,,\nonumber\\
 	 &&\Phi_J(z)=(R/z)^{J-3/2} e^{-\varphi(z)/2}\phi_J(\zeta) \,,\nonumber\\
&&(m_5R)^2 =-(2-J)^2 + L^2 \,,
 \label{mapping}
 \ee
 where $z,R$ and $m_5$ are the fifth dimension, radius of curvature and mass parameter in $\mathrm{AdS}_5$ spacetime. The confinement potential in physical spacetime is then fixed by the form of the dilaton field in $\mathrm{AdS}_5$:
 \begin{equation}
	U(\zeta, J)= \frac{1}{2} \varphi^{\prime\prime}(z) + \frac{1}{4} \varphi^{\prime}(z)^2 + \left(\frac{2J-3}{2 z} \right)\varphi^{\prime} (z) \,.
\label{dilation-U}
\end{equation}
The holographic mapping does not fix $\varphi(z)$ but the underlying conformal symmetry requires it to be quadratic: $\varphi(z)=\kappa^2 z^2$ \cite{Brodsky:2013ar}. The confining potential in Eq.~(\ref{hSE-mesons}) is then given by 
\begin{equation}
	U(\zeta, J)=\kappa^4 \zeta^2 + 2 \kappa^2(J-1) \;.
	\label{unique-U}
\end{equation}
 We refer to the emerging mass scale, $\kappa$, which simultaneously sets the strength of the dilaton field in $\mathrm{AdS}_5$ and the confinement scale in physical spacetime, as the light-front holographic mass scale. Solving Eq.~(\ref{hSE-mesons}) with the potential, Eq.~(\ref{unique-U}), yields
\be
 	\phi_{nL}(\zeta)= \kappa^{1+L} \sqrt{\frac{2 n !}{(n+L)!}} \zeta^{1/2+L}\exp{\left(\frac{-\kappa^2 \zeta^2}{2}\right)}  ~ L_n^L(\kappa^2 \zeta^2) \,,
 \label{phi-zeta}
 \ee
so that, for $n=J=L=0$ states, Eq.~(\ref{holographic-wfn-impact}) becomes
 \begin{equation}
 	\Psi(x,\zeta)=\frac{\kappa}{\sqrt{\pi}} X(x) \exp\left(-\frac{\kappa^2 \zeta^2}{2}\right) \;.
 	\label{pionwf}
 \end{equation}

 The longitudinal wavefunction, $X(x)$, can be found by mapping the space-like EM form factor in physical spacetime onto the EM form factor in $\mathrm{AdS}_5$. The latter is computed using the free EM current propagator i.e. the equation of motion for the EM field in pure (without a dilaton field) $\mathrm{AdS}_5$. 
 In physical spacetime, the EM form factor is given by the Drell-Yan-West formula \cite{Drell:1969km,West:1970av}:
  \be
 	F_{\mathrm{DYW}}(Q^2)= 2 \pi \int_0^1 \frac{\mathrm{d} x}{x\bar{x}} \int \zeta \mathrm{d} \zeta J_0 \left(\zeta Q \sqrt{\frac{\bar{x}}{x}}\right) |\Psi(x, \zeta)|^2 \;,
 \label{DYW-FF}
 \ee
 while, in $\mathrm{AdS}_5$, it is given by
 \begin{equation}
 	F_{\mathrm{AdS}}(Q^2)= R^3 \int_0^1 \mathrm{d} x \int \frac{\mathrm{d} z}{z^3} J_0\left( z Q \sqrt{\frac{\bar{x}}{x}} \right) |\Phi(z)|^2\;.
 \label{AdS-FF}
 \end{equation}
The holographic mapping of Eq.~(\ref{DYW-FF}) onto Eq.~(\ref{AdS-FF}) requires that $X(x)=\sqrt{x\bar{x}}$, so that  
 \begin{equation}
 	\Psi(x,\zeta)=\frac{\kappa}{\sqrt{\pi}} \sqrt{x\bar{x}}\, \exp\left(-\frac{\kappa^2 \zeta^2}{2}\right) \;.
 	\label{universal-wf}
 \end{equation}
After accounting for light quark masses using the Brodsky-de T\'eramond prescription \cite{Brodsky:2008pg}, Eq.~(\ref{universal-wf}) becomes
 \be
 	\Psi(x,\zeta) \propto \sqrt{x\bar{x}} \,\exp\left(-\frac{\kappa^2 \zeta^2}{2}\right) \, \exp \left(-\frac{\bar{x} m^2_{1} + x m^2_{2}}{2\kappa^2x\bar{x}}\right)  \;,
 	\label{universal-wf-quark-masses}
 \ee
 where $m_1$ is the quark mass. In a meson, $m_2$ is the antiquark mass and, in a baryon, it is the diquark mass.

Alternatively, the $\mathrm{AdS}_5$ EM form factor can be computed using a dressed EM current propagator, i.e. the equation of motion for an EM field in a dilation-modified $\mathrm{AdS}_5$. In this case, the EM form factor admits a twist expansion, and to leading twist accuracy, is given by \cite{Brodsky:2014yha}
\begin{equation}
	F_{\mathrm{AdS}}(Q^2)=\int \mathrm{d} x x^{Q^2/4\kappa^2} \;.
\label{AdS-FF-dilaton}
\end{equation}
We can recover Eq.~(\ref{AdS-FF-dilaton}) in physical spacetime by using the wavefunction,
\begin{equation}
  	\Psi(x, \zeta)=\frac{\kappa\bar{x}}{\sqrt{\pi} \ln(1/x)} \exp \left(-\frac{\kappa^2 \zeta^2\bar{x}}{2x\ln (1/x)} \right) 
  	\label{effective-wf}
  \end{equation}
in the Drell-Yan-West formula, Eq.~(\ref{DYW-FF}).  

Unlike Eq.~(\ref{universal-wf}), Eq.~(\ref{effective-wf}) is not symmetric under the transformation $x \leftrightarrow 1-x$, and can be thought as an effective wavefunction, in contradistinction with Eq.~(\ref{universal-wf}) which is a true valence wavefunction. In this paper, we shall use Eq.~(\ref{universal-wf}) and not Eq.~(\ref{effective-wf}). It is thus clear that we are not assuming any contributions from higher Fock sectors. This distinguishes our paper from previous work \cite{Maji:2016yqo,Gutsche:2014yea,Mondal:2015uha} which use Eq.~(\ref{effective-wf}). 
 
Eq.~(\ref{phi-zeta}) implies that states with the same $(n,L)$ quantum numbers share the same holographic wavefunction, regardless of the value of $J$. For instance, the pion ($J=0$) and the $\rho~(J=1)$ meson, have identical holographic wavefunctions and thus identical decay constants \cite{Ahmady:2020mht}. This is a direct consequence of the decoupling of the quark helicity from the confinement dynamics in light-front holography. To overcome this shortcoming, we account for the helicities of quarks in a dynamical way, i.e. by multiplying the universal holographic wavefunction by a momentum-dependent helicity wavefunction, so that
 \begin{equation}
 	\Psi_{h_1 h_2}(x,\mathbf{k})= S_{h_1 h_2}(x, \mathbf{k}) \times \Psi(x, k_\perp)\,,
 \label{spin-improved-wf}
 \end{equation} 
 where $\mathbf{k}=k_\perp e^{i\theta_{k_\perp}}$ is conjugate to $\mathbf{b}$, and $\Psi(x,k_\perp)$ is the two-dimensional Fourier transform of Eq.~(\ref{universal-wf-quark-masses}). The indices $h_{1,2}$ denote the helicities of $q_{1,2}$ (recall that for mesons, $q_1 \equiv q, q_2 \equiv \bar{q}$ while for baryons, $q_1 \equiv q, q_2 \equiv [qq]$).  We assume that $S_{h_1 h_2}(x,\mathbf{k})$ follows from the point-like coupling of the hadron with the partons, while all bound states effects are captured by the holographic wavefunction $\Psi(x, k_\perp)$. For vector mesons, we thus have \cite{Forshaw:2012im,Ahmady:2016ujw}
\be
	S^{V(\lambda)}_{h_q h_{\bar{q}}}(x,\mathbf{k}) \propto \frac{\bar{v}_{h_{\bar{q}}}((1-x)P^+, -\mathbf{k})}{\sqrt{\bar{x}}}[\epsilon^{\lambda}_{V} \cdot \gamma] \frac{u_{h_q}(xP^+,\mathbf{k})}{\sqrt{x}} \,,
\label{vector-spin}
\ee
where $\epsilon^{\lambda}_V$ are the polarization $4$-vectors for the vector meson, while for pseudoscalar mesons, 
\begin{equation}
 S^P_{h_q h_{\bar{q}}}(x,\mathbf{k}) \propto	\frac{\bar{v}_{h_{\bar{q}}}(\bar{x}P^+, -\mathbf{k})}{\sqrt{\bar{x}}}   [\gamma^5]  \frac{u_{h_q}(xP^+,\mathbf{k})}{\sqrt{x}} \;.
 \label{pseudoscalar-spin}  
 \end{equation} 
Indeed, it was recently shown in Ref. \cite{Ahmady:2020mht}, that Eqs.~(\ref{vector-spin}) and (\ref{pseudoscalar-spin}), together with the universal holographic wavefunction, Eq.~(\ref{universal-wf-quark-masses}), lead to successful predictions for the vector-to-pseudoscalar meson radiative form factors.

In this paper, we extend this idea to the nucleons. Keeping in mind that Eq.~(\ref{hSE-mesons}) is also true for a quark-diquark bound state, we assume that the spin structure follows from the point-like coupling of a nucleon to a quark and a diquark. Now, the spin of the diquark can be $S_D=0$ or $S_D=1$, so that we have two possible spin structures. With a scalar $(S_D=0)$ diquark, we choose
 \begin{equation}
 	S^N_{h_N h_q}(x, \mathbf{k}) \propto \frac{\bar{u}_{h_q}(xP^+,\mathbf{k})}{\sqrt{x}} [{1}] \frac{u_{h_N}(P^+,\mathbf{0})}{\sqrt{1}}
 \label{scalar-diquark}
 \end{equation}
 analogous to the spin structure, Eq.~(\ref{pseudoscalar-spin}), for the pseudoscalar mesons. For an axial-vector diquark $(S_D=1)$, we assume
   \begin{equation}
   	S^{N(\lambda)}_{h_N h_q}(x, \mathbf{k}) \propto \frac{\bar{u}_{h_q}(xP^+,\mathbf{k})}{\sqrt{x}} [(\epsilon^{\lambda}_{D} \cdot \gamma) \gamma^5 ] \frac{u_{h_N}(P^+,\mathbf{0})}{\sqrt{1}}  
  \label{vector-diquark}
  \end{equation}
 analogous to the spin structure, Eq.~(\ref{vector-spin}), for a vector meson. In Eq.~(\ref{vector-diquark}), the diquark polarization vectors read 
\be
\epsilon^0_D &= & \left(\frac{(1-x)P^+}{M_D}, -\frac{M_D}{\bar{x}P^+}, 0_\perp \right)\,,\\
{\epsilon^\pm_D} &=&\frac{1}{\sqrt{2}} (0,0,\mp 1,-i)\,,
\ee
where $M_D$ is the diquark mass. Note that the spin structures, given by Eqs.~ (\ref{scalar-diquark}) and (\ref{vector-diquark}), were proposed in Ref. \cite{Brodsky:2000ii} (for scalar diquarks only) and in Ref. \cite{Ellis:2008in} (for both scalar and axial-vector diquarks) with the assumption of a heavy diquark, i.e. $M_D=M_N$ and $1-x \approx 1$. Here, we do not make this approximation, but we do recover the results of Ref. \cite{Ellis:2008in} when doing so.

 In light-front holographic QCD, the EM form factors are directly computed using a dressed EM current in $\mathrm{AdS}_5$ \cite{Chakrabarti:2013dda,Sufian:2016hwn,Gutsche:2012bp}.   
The alternative approach, which we follow here, is to compute the form factors in physical spacetime using the DYW formula, Eq.~(\ref{DYW-FF}), with a wavefunction which is itself obtained from light-front holography. The virtue of this approach is that, once the physical spacetime nucleon holographic wavefunction is known, other QCD distributions like  the Transverse Momentum Dependent Parton Distributions (TMDs) and Generalized Parton Distributions (GPDs), etc., can be predicted. On the other hand, the computation of the EM form factor is then restricted to the space-like region. This approach is taken in \cite{Maji:2016yqo,Gutsche:2014yea,Mondal:2015uha,Gutsche:2013zia} using  the spin wavefunctions of Ref. \cite{Ellis:2008in} together with the effective holographic wavefunction, Eq.~(\ref{effective-wf}), or models inspired by Eq.~(\ref{effective-wf}).  Refs.  \cite{Gutsche:2014yea,Mondal:2015uha,Gutsche:2013zia} consider only a scalar diquark while Ref. \cite{Maji:2016yqo} considers both a scalar and axial-vector diquark. The number of adjustable parameters in Refs. \cite{Maji:2016yqo,Mondal:2015uha,Gutsche:2013zia} is typically large $(>10)$.  
   
We shall use Eq.~(\ref{scalar-diquark}) and (\ref{vector-diquark}), together with the universal holographic wavefunction, Eq. (\ref{universal-wf-quark-masses}), to predict the form factors of the nucleons. Our goal in this paper is two-fold. First, we improve upon existing calculations using holographic light-front wavefunctions \cite{Maji:2016yqo,Gutsche:2014yea,Gutsche:2013zia} by using much fewer free parameters: our only adjustable parameters are the quark and diquark masses. Once fixed, using a judiciously chosen data set, our remaining predictions do not contain any adjustable parameters. We also predict the charge and magnetic radii of the nucleons, as well as the axial charge of the proton. 
Second, this paper, together with previous work \cite{Forshaw:2012im,Ahmady:2016ujw,Ahmady:2016ufq,Ahmady:2019yvo,Kaur:2020emh,Ahmady:2020mht}, demonstrates that
nucleons (quark-diquark) and mesons (quark-antiquark) share the same universal two-parton holographic
wavefunction. 
%{\color{blue}Second, this paper, together with previous work \cite{Forshaw:2012im,Ahmady:2016ujw,Ahmady:2016ufq,Ahmady:2019yvo,Ahmady:2020mht}, demonstrate that within the two particle approximation, the light mesons (quark-antiquark) and the nucleons (quark-diquark) can be treated in a unified framework, where they share the same `\textit{form}' of the two particle  light-front holographic wavefunction and their helicity-dependent parts of the wavefunctions are modified differently by their spin structures.}

Note that the nucleon EM form factors can be calculated  in model-independent ways e.g., in effective field theories \cite{Kubis:2000zd} and lattice QCD \cite{Green:2014xba} and it is not our goal here to compare or compete with these calculations.

%%%%%%%%%%%%%%%%%%%%%%%%%%%%%%%%%%%%%%%%
\section{Spin-improved holographic wavefunctions for the nucleons}

Using light-front field theory \cite{Lepage:1980fj}, we start by deriving explicit expressions for the spin-improved holographic wavefunctions of the nucleons using Eq.~(\ref{spin-improved-wf}), together with the spin structures given by Eq.~(\ref{scalar-diquark}) and Eq.~(\ref{vector-diquark}). Our notation for the resulting wavefunction is: $\Psi^{h_N [D]}_{h_q, h_D}$ where $D=\{S, A\}$ denotes a scalar or axial-vector diquark, while $h_N=\{\uparrow,\downarrow\}$, $h_q=\{+1/2,-1/2\}$ and $h_D=\{0,\pm 1\}$ are the helicities of the nucleon, quark and diquark respectively. 

For a scalar diquark, we obtain
\be
\label{scalar-wfs} 
\Psi^{\uparrow[S]}_{\frac{1}{2},0}(x, \mathbf{k})&=&  \left(M_N+\frac{m_q}{x}\right)\Psi(x, k_\perp)\,, \nonumber\\
\Psi^{\uparrow[S]}_{-\frac{1}{2},0}(x, \mathbf{k})&=& - \left(\frac{k_\perp e^{i\theta_{k_\perp}}}{x}\right)\Psi(x, k_\perp)\,, \nonumber\\
\Psi^{\downarrow[S]}_{\frac{1}{2},0}(x, \mathbf{k})&=&  \left(\frac{k_\perp e^{-i\theta_{k_\perp}}}{x}\right)\Psi(x, k_\perp)\,, \nonumber\\
\Psi^{\downarrow[S]}_{-\frac{1}{2},0}(x,\mathbf{k})&=&  \left(M_N+\frac{m_q}{x}\right)\Psi(x,k_\perp) \,,
\ee
where $M_N$ is the nucleon mass and $m_q$ is the mass of the quark. For an axial-vector diquark, we find
\be
\label{vector-wfs-up}
\Psi^{\uparrow[A]}_{\frac{1}{2},0}(x,\mathbf{k})&=&  -\left(\frac{M_N m_q}{M_D}\frac{\bar{x}}{x}+\frac{M_D}{\bar{x}}\right)\Psi(x,k_\perp)\,, \nonumber\\
\Psi^{\uparrow[A]}_{-\frac{1}{2},0}(x, \mathbf{k})&=& \left(\frac{k_\perp e^{i\theta_{k_\perp}}}{x}\frac{\bar{x}M_N}{M_D} \right) \Psi(x,k_\perp)\,, \nonumber\\
\Psi^{\uparrow[A]}_{\frac{1}{2},1}(x, \mathbf{k})&=&  \left(\frac{k_\perp e^{-i\theta_{k_\perp}}}{x}\right)\Psi(x,k_\perp)\,, \nonumber\\
\Psi^{\uparrow[A]}_{-\frac{1}{2},1}(x, \mathbf{k})&=&  \left(M_N+\frac{m_q}{x}\right)\Psi(x,k_\perp)\,, \nonumber\\
\Psi^{\uparrow[A]}_{\frac{1}{2},-1}(x,\mathbf{k})&=& 0 \,, \nonumber\\
\Psi^{\uparrow[A]}_{-\frac{1}{2},-1}(x,\mathbf{k})&=& 0\,, 
\ee
for a spin-up nucleon, and
\be
\label{vector-wfs-down}
\Psi^{\downarrow[A]}_{\frac{1}{2},0}(x, \mathbf{k})&=&  \left(\frac{k_\perp e^{-i\theta_{k_\perp}}}{x}\frac{(\bar{x})M_N}{M_D} \right) \Psi(x,k_\perp)\,, \nonumber\\
\Psi^{\downarrow[A]}_{-\frac{1}{2},0}(x, \mathbf{k})&=&  \left(\frac{M_N m_q}{M_D}\frac{\bar{x}}{x}+\frac{M_D}{\bar{x}}\right)\Psi(x, k_\perp)\,, \nonumber\\
\Psi^{\downarrow[A]}_{\frac{1}{2},1}(x, \mathbf{k})&=&  0\,, \nonumber\\
\Psi^{\downarrow[A]}_{-\frac{1}{2},1}(x, \mathbf{k})&=& 0\,,\nonumber\\
\Psi^{\downarrow[A]}_{\frac{1}{2},-1}(x, \mathbf{k})&=&  - \left(M_N+\frac{m_q}{x}\right)\Psi(x,k_\perp)\,,   \nonumber\\
\Psi^{\downarrow[A]}_{-\frac{1}{2},-1}(x, \mathbf{k})&=& \left(\frac{k_\perp e^{i\theta_{k_\perp}}}{x} \right)\Psi(x,k_\perp)\,, 
\ee
for a spin-down nucleon. 

Using these wavefunctions, the nucleon state can be expanded on a quark-diquark basis. For example, for a spin-up proton state, we have
\be
\label{proton-up}
\mid P\uparrow\rangle &=& \int \frac{\mathrm{d}x \mathrm{d}^2 \mathbf{k}} {16\pi^3 \sqrt{x\bar{x}}}  \Bigg[ \Psi_{\frac{1}{2},0}^{\uparrow[S]}(x, \mathbf{k}) \mid [ud]_S u^\uparrow\rangle  + \Psi_{-\frac{1}{2},0}^{\uparrow[S]}(x, \mathbf{k}) \mid [ud]_S  u^\downarrow \rangle\nonumber \\
&&+ \Psi_{\frac{1}{2},0}^{\uparrow[A]}(x, \mathbf{k}) \left( \frac{1}{3}\mid [ud]^{0}_A u^\uparrow\rangle - \frac{\sqrt{2}}{3}  \mid [uu]^{0}_A d^\uparrow\rangle \right)\nonumber\\
&& + \Psi_{-\frac{1}{2},0}^{\uparrow[A]}(x, \mathbf{k}) \left(- \frac{1}{3}\mid [ud]^{0}_A u^\downarrow\rangle + \frac{\sqrt{2}}{3}  \mid [uu]^{0}_A d^\downarrow\rangle \right)\nonumber\\
&& +  \Psi_{\frac{1}{2},-1}^{\uparrow[A]}(x, \mathbf{k}) \left( \frac{\sqrt{2}}{3}\mid [ud]^{-1}_A u^\uparrow\rangle - \frac{2}{3}  \mid [uu]^{-1}_A d^\uparrow\rangle \right)\\
&&  +  \Psi_{-\frac{1}{2},1}^{\uparrow[A]}(x, \mathbf{k}) \left( -\frac{\sqrt{2}}{3}\mid [ud]^{1}_A u^\downarrow\rangle + \frac{2}{3}  \mid [uu]^{1}_A d^\downarrow\rangle \right) \nonumber\\
&&  +   \Psi_{\frac{1}{2},1}^{\uparrow[A]}(x, \mathbf{k}) \left( -\frac{\sqrt{2}}{3}\mid [ud]^{1}_A u^\uparrow\rangle + \frac{2}{3}  \mid [uu]^{1}_A d^\uparrow\rangle \right) \nonumber\\
&&  +   \Psi_{-\frac{1}{2},-1}^{\uparrow[A]}(x, \mathbf{k}) \left( -\frac{\sqrt{2}}{3}\mid [ud]^{-1}_A u^\downarrow\rangle + \frac{2}{3}  \mid [uu]^{-1}_A d^\downarrow\rangle \right) \Bigg]\nonumber
\ee
where $[qq]^{h_D}_{D}$ indicates a diquark. It is worth making several comments regarding Eq.~(\ref{proton-up}). 

First, if we neglect the axial-vector diquark contributions, Eq.~(\ref{proton-up}) implies that the proton is a superposition of $L=0$ and $L=1$ states. This is also predicted in the supersymmetric formulation of light-front holography \cite{Brodsky:2016rvj} where baryons and mesons, differing by one unit of orbital angular momentum, are supersymmetric partners. Tetraquarks are the second supersymmetric partners of baryons with the same orbital angular momentum. For example, the proton has superpartners $h_1(1170) (L=1)$ and $\sigma(500) (L=0)$,  the latter being identified as a tetraquark candidate \cite{Nielsen:2018uyn}. 
  However, unlike the supersymmetric holographic wavefunction, our wavefunction remains asymmetric under the interchange $x \leftrightarrow \bar{x}$, even in the chiral limit. This is because our the spin wavefunction does not decouple from the dynamics, as is the case in light-front holography.

Second, the spin-statistics theorem dictates that the diquark cluster must have isospin and spin either both equal to zero or both equal to one, so that the combined spin-isospin wavefunction is symmetric. This is because the relative orbital angular momentum of the two quarks is zero and their colour wavefunction is antisymmetric. Therefore the spin-statistics theorem does not allow a $[uu]$ scalar diquark, implying that the $d$-quark cannot be single.  Therefore, in a purely quark-scalar-diquark model for the proton, the $d$-quark cannot be active. To remedy this shortcoming, we also consider the possibility of an axial-vector diquark. Then, as can be seen in Eq.~(\ref{proton-up}), this opens up the possibility that the $d$-quark to be active. 

Thirdly, in the absence of dynamical spin effects, i.e. when all (non-vanishing) $\Psi^{h_p[D]}_{h_q, h_D}(x,\mathbf{k}) \to \Psi(x, k_\perp)$, Eq.~(\ref{proton-up}) reduces to the spin-flavour $\mbox{SU}(4)$ quark model expansion for the proton state \cite{Ma:2002ir}. Thus, we can think of dynamical spin effects as breaking the $\mbox{SU}(4)$ spin-flavour symmetry of the quark model. 

 Finally, interchanging $\uparrow$ and $\downarrow$  in Eq.~(\ref{proton-up}) yields the expansion for a spin-down proton, and the isospin transformation $u \leftrightarrow d$ yields the expansion for the neutron state.

\section{Analytic expressions for the form factors}

At the quark level, the Dirac and Pauli form factors, $F^f_1(q^2)$ and $F^f_2(q^2)$, are defined via
\be 
\langle P+q,\uparrow|J_{\mathrm{EM},f}^{+}(0)|P,\uparrow \rangle &=& 2P^+ F^f_1(q^2),
\label{DiracFF}
\ee
\be
\langle P+q,\uparrow|J_{\mathrm{EM},f}^{+}(0)|P,\downarrow\rangle = -(q^1-iq^2)\frac{P^+}{M_N}F^f_2(q^2),
\label{PauliFF}
\ee 
where 
\begin{equation}
	J^{\mu}_{\mathrm{EM},f}=\bar{\psi}_f\gamma^\mu\psi_f
\end{equation}
is the EM vector quark current with flavour $f=[u,d]$. On the other hand, the axial form factor, $G^f_A(q^2)$, parametrizes the nucleon matrix elements of the axial EW current: 
\be
\langle P+q,\uparrow|J_{\mathrm{EW},f}^{+}(0)|P,\uparrow \rangle &=& 2P^+ G^f_A(q^2)\,,
\label{AxialFF}
\ee
where 
\begin{equation}
	J_{\mathrm{EW},f}^{\mu}=\bar{\psi}_f\gamma^\mu\gamma^5\psi_f \;.
\end{equation}

To evaluate the left-hand-sides of  Eqs.~(\ref{DiracFF}),  (\ref{PauliFF}) and (\ref{AxialFF}), we expand the nucleon state using Eq.~(\ref{proton-up}).  As we mentioned above, we compute the form factors in the impulse approximation where the single quark in Eq.~(\ref{proton-up}) is the active quark coupling to the photon (or $Z$) while the diquark cluster is spectator. 

The scalar diquark contributions to the EM form factors then read
 \be 
F^{S(u)}_1(Q^2)=\int \mathrm{d}x \frac{\mathrm{d}^2 \mathbf{k}}{16\pi^3} \left[{\psi^{S\uparrow}_{\frac{1}{2},0}}^\dagger (x,\mathbf{k}^\prime)  \psi^{S\uparrow}_{\frac{1}{2},0}(x, \mathbf{k}) +\,{\psi^{S\uparrow}_{-\frac{1}{2},0}}^\dagger (x, \mathbf{k}^\prime)  \psi^{S\uparrow}_{-\frac{1}{2},0}(x, \mathbf{k}) \right]\,,
\ee 
and
 \be
 F^{S(u)}_2(Q^2)=\, -\frac{2M_N}{q_\perp e^{-i\theta_{q_\perp}}} \int \mathrm{d}x \frac{\mathrm{d}^2 \mathbf{k}}{16\pi^3} \left[{\psi^{S\uparrow}_{\frac{1}{2},0}}^\dagger (x, \mathbf{k}^\prime)  \psi^{S\downarrow}_{\frac{1}{2},0}(x, \mathbf{k})  +\,{\psi^{S\uparrow}_{-\frac{1}{2},0}}^\dagger (x, \mathbf{k}^\prime)  \psi^{S\downarrow}_{-\frac{1}{2},0}(x, \mathbf{k}) \right]\,,
\ee
where $\mathbf{k}^\prime=\mathbf{k} - \bar{x} \mathbf{q}$. Using our spin-improved holographic wavefunctions, Eq.~(\ref{scalar-wfs}), we find
 \begin{align}
 F^{S(u)}_1(Q^2)  \propto & \int \mathrm{d}x \left[ \left(M_N+\frac{m_u}{x}\right)^2+\frac{(1-x)\kappa^2}{x}\left(1-\frac{(1-x)q_\perp^2}{4\kappa^2 x}\right)\right]\nonumber \\& \times\exp\left(-\frac{(1-x)q_\perp^2}{4\kappa^2 x}\right) \exp\left(-\frac{xM_{ud}^2+(1-x)m_u^2}{\kappa^2 x(1-x)}\right), 
\end{align}
and
\be
F^{S(u)}_2(Q^2)\propto   \int \mathrm{d}x \left[ \left(M_N+\frac{m_u}{x}\right)\frac{\bar{x}}{x}\right]\exp\left(-\frac{\bar{x}q_\perp^2}{4\kappa^2 x}\right) \exp\left(-\frac{xM_{ud}^2+\bar{x} m_u^2}{\kappa^2 x\bar{x}}\right) \;.
\ee

We next proceed with the axial-vector diquark contributions to the EM form factors: 
 \begin{align}
 F^{A(u)}_1(Q^2) =& \int \mathrm{d}x\, \frac{\mathrm{d}^2 \mathbf{k}}{16\pi^3} \,
 \Bigg[ \frac{1}{9} \left({\psi^{A\uparrow}_{\frac{1}{2},0}}^\dagger (x, \mathbf{k}^\prime)  \psi^{A\uparrow}_{\frac{1}{2},0}(x, \mathbf{k}) +
{\psi^{A\uparrow}_{-\frac{1}{2},0}}^\dagger (x, \mathbf{k}^\prime)  \psi^{A\uparrow}_{-\frac{1}{2},0}(x, \mathbf{k})\right) \nonumber\\
&+\frac{2}{9}\left(
{\psi^{A\uparrow}_{\frac{1}{2},1}}^\dagger (x, \mathbf{k}^\prime)  \psi^{A\uparrow}_{\frac{1}{2},1}(x, \mathbf{k}) +\,{\psi^{A\uparrow}_{-\frac{1}{2},1}}^\dagger (x, \mathbf{k}^\prime)  \psi^{A\uparrow}_{-\frac{1}{2},1}(x, \mathbf{k})\right) \Bigg] \;,
\end{align}
 \begin{align}
 F^{A(d)}_1(Q^2)=& \int \mathrm{d}x\, \frac{\mathrm{d}^2 \mathbf{k}}{16\pi^3} \,
 \Bigg[ \frac{2}{9} \left({\psi^{A\uparrow}_{\frac{1}{2},0}}^\dagger (x, \mathbf{k}^\prime)  \psi^{A\uparrow}_{\frac{1}{2},0}(x, \mathbf{k}) +
{\psi^{A\uparrow}_{-\frac{1}{2},0}}^\dagger (x, \mathbf{k}^\prime)  \psi^{A\uparrow}_{-\frac{1}{2},0}(x, \mathbf{k})\right) \nonumber\\
&+\frac{4}{9}\left(
{\psi^{A\uparrow}_{\frac{1}{2},1}}^\dagger (x, \mathbf{k}^\prime)  \psi^{A\uparrow}_{\frac{1}{2},1}(x, \mathbf{k}) + \,{\psi^{A\uparrow}_{-\frac{1}{2},1}}^\dagger (x, \mathbf{k}^\prime)  \psi^{A\uparrow}_{-\frac{1}{2},1}(x, \mathbf{k})\right) \Bigg] \;,
\end{align}
 and
 \begin{align}
F^{A(u)}_2(Q^2)  =-\frac{2M_N}{q^1-iq^2}\int \mathrm{d}x\, \frac{\mathrm{d}^2\mathbf{k}}{16\pi^3}\,\frac{1}{9}
\Big[{\psi^{A\uparrow}_{\frac{1}{2},0}}^\dagger(x, \mathbf{k}^\prime)\psi^{A\downarrow}_{\frac{1}{2},0} (x, \mathbf{k}) +\,{\psi^{A\uparrow}_{-\frac{1}{2},0}}^\dagger (x, \mathbf{k}^\prime)\psi^{A\downarrow}_{-\frac{1}{2},0} (x, \mathbf{k})\Big]\,,
\end{align}
\be
F^{A(d)}_2(Q^2)  =-\frac{2M_N}{q^1-iq^2}\int \mathrm{d}x\, \frac{\mathrm{d}^2\mathbf{k}}{16\pi^3}\,\frac{2}{9}
\Big[{\psi^{A\uparrow}_{\frac{1}{2},0}}^\dagger(x, \mathbf{k}^\prime)\psi^{A\downarrow}_{\frac{1}{2},0} (x, \mathbf{k}) +\,{\psi^{A\uparrow}_{-\frac{1}{2},0}}^\dagger (x, \mathbf{k}^\prime)\psi^{A\downarrow}_{-\frac{1}{2},0} (x, \mathbf{k})\Big]\,.
\ee
 Using our spin-improved wavefunctions, namely Eqs. (\ref{vector-wfs-down}) and (\ref{vector-wfs-up}), we find that
\begin{align}
 F_1^{A(u)}(Q^2) \propto &\int \mathrm{d}x\, \Bigg[\frac{1}{9} \left\{ \left(\frac{M_N m_u}{M_{ud}} \frac{\bar{x}}{x}+\frac{M_{ud}}{\bar{x}}\right)^2  +\,\kappa^2 x \bar{x} \left(\frac{M_N\bar{x}}{M_{ud} x}\right)^2\left(1-\frac{q_\perp^2\bar{x}}{4\kappa^2 x}\right)\right\}  \nonumber \\
& +\, \frac{2}{9} \left\{\left(M_N+\frac{m_u}{x}\right)^2+\kappa^2\frac{\bar{x}}{x} \left(1-\frac{q_\perp^2\bar{x}}{4\kappa^2 x}\right)\right\}\Bigg]  \exp\left(-\frac{\bar{x}q_\perp^2}{4\kappa^2 x}\right) \exp\left(-\frac{xM_{ud}^2+\bar{x}m_u^2}{\kappa^2 x\bar{x}}\right) \;,
\end{align}
\begin{align}
 F_1^{A(d)}(Q^2)  \propto &\int \mathrm{d}x\, \Bigg[\frac{2}{9} \left\{ \left(\frac{M_N m_u}{M_{uu}} \frac{\bar{x}}{x}+\frac{M_{uu}}{\bar{x}}\right)^2 +\,\kappa^2 x \bar{x} \left(\frac{M_N\bar{x}}{M_{uu} x}\right)^2\left(1-\frac{q_\perp^2\bar{x}}{4\kappa^2 x}\right)\right\}  \nonumber \\
&  +\, \frac{4}{9} \left\{\left(M_N+\frac{m_u}{x}\right)^2+\kappa^2\frac{\bar{x}}{x} \left(1-\frac{q_\perp^2\bar{x}}{4\kappa^2 x}\right)\right\}\Bigg]  \exp\left(-\frac{\bar{x}q_\perp^2}{4\kappa^2 x}\right) \exp\left(-\frac{xM_{uu}^2+\bar{x}m_u^2}{\kappa^2 x\bar{x}}\right) \;,
\end{align}
 and
\be
F^{A(u)}_2(Q^2)\propto \int \mathrm{d}x\, \frac{1}{9}\,\frac{\bar{x}^2}{x}\left(\frac{M_N m_u}{M_{ud}} \frac{\bar{x}}{x}+\frac{M_{ud}}{1-x}\right)\frac{M_N}{M_{ud}}\, \exp\left(-\frac{\bar{x}q_\perp^2}{4\kappa^2 x}\right)
\exp\left(-\frac{xM_{ud}^2+\bar{x}m_u^2}{\kappa^2 x\bar{x}}\right) \;,
\ee
\be
F^{A(d)}_2(Q^2)\propto\int \mathrm{d}x\, \frac{2}{9}\,\frac{\bar{x}^2}{x}\left(\frac{M_N m_d}{M_{uu}} \frac{\bar{x}}{x}+\frac{M_{uu}}{\bar{x}}\right)\frac{M_N}{M_{uu}} \,\exp\left(-\frac{\bar{x}q_\perp^2}{4\kappa^2 x}\right) \exp\left(-\frac{xM_{uu}^2+\bar{x} m_d^2}{\kappa^2 x\bar{x}}\right) \;.
\ee
Finally, we find the scalar and axial-vector diquark contributions to the axial form factor:
    \be 
G^{S(u)}_A(Q^2)=\int \mathrm{d}x \,\frac{\mathrm{d}^2 \mathbf{k}}{16\pi^3}\, \left[{\psi^{S\uparrow}_{\frac{1}{2},0}}^\dagger (x, \mathbf{k}^\prime)  \psi^{S\uparrow}_{\frac{1}{2},0}(x, \mathbf{k})
-\,{\psi^{S\uparrow}_{-\frac{1}{2},0}}^\dagger (x, \mathbf{k}^\prime)  \psi^{S\uparrow}_{-\frac{1}{2},0}(x, \mathbf{k}) \right]\,,
\ee
\begin{align}
 G^{A(u)}_A(Q^2)=& \int \mathrm{d}x\, \frac{\mathrm{d}^2 \mathbf{k}}{16\pi^3}\, 
 \Bigg[ \frac{1}{9} \left({\psi^{A\uparrow}_{\frac{1}{2},0}}^\dagger (x, \mathbf{k}^\prime)  \psi^{A\uparrow}_{\frac{1}{2},0}(x, \mathbf{k}) -
{\psi^{A\uparrow}_{-\frac{1}{2},0}}^\dagger (x, \mathbf{k}^\prime)  \psi^{A\uparrow}_{-\frac{1}{2},0}(x, \mathbf{k})\right)\nonumber\\
& +\frac{2}{9}\left(
{\psi^{A\uparrow}_{\frac{1}{2},1}}^\dagger (x, \mathbf{k}^\prime)  \psi^{A\uparrow}_{\frac{1}{2},1}(x, \mathbf{k}) - {\psi^{A\uparrow}_{-\frac{1}{2},1}}^\dagger (x, \mathbf{k}^\prime)  \psi^{A\uparrow}_{-\frac{1}{2},1}(x, \mathbf{k})\right) \Bigg] \;,
\end{align}
 and
\begin{align}
 G^{A(d)}_A(Q^2)=& \int \mathrm{d}x\, \frac{\mathrm{d}^2 \mathbf{k}}{16\pi^3}\, 
 \Bigg[ \frac{2}{9} \left({\psi^{A\uparrow}_{\frac{1}{2},0}}^\dagger (x, \mathbf{k}^\prime)  \psi^{A\uparrow}_{\frac{1}{2},0}(x, \mathbf{k}) -
{\psi^{A\uparrow}_{-\frac{1}{2},0}}^\dagger (x, \mathbf{k}^\prime)  \psi^{A\uparrow}_{-\frac{1}{2},0}(x, \mathbf{k})\right) \nonumber\\
&+\frac{4}{9}\left(
{\psi^{A\uparrow}_{\frac{1}{2},1}}^\dagger (x, \mathbf{k}^\prime)  \psi^{A\uparrow}_{\frac{1}{2},1}(x, \mathbf{k}) -{\psi^{A\uparrow}_{-\frac{1}{2},1}}^\dagger (x, \mathbf{k}^\prime)  \psi^{A\uparrow}_{-\frac{1}{2},1}(x, \mathbf{k})\right) \Bigg] \;,
\end{align}
 Using our spin-improved wavefunctions, given by Eqs. (\ref{scalar-wfs}) and (\ref{vector-wfs-up}), we find that  
\begin{align}
G^{S(u)}_A(Q^2)\propto  
& \int dx \,\left[ \left(M_N+\frac{m_u}{x}\right)^2-\frac{(1-x)\kappa^2}{x}\left(1-\frac{(1-x)q_\perp^2}{4\kappa^2 x}\right)\right] \nonumber \\
&\times \exp\left(-\frac{(1-x)q_\perp^2}{4\kappa^2 x}\right) \exp\left(-\frac{xM_{ud}^2+(1-x)m_u^2}{\kappa^2 x(1-x)}\right),
\end{align}
\begin{align}
 G^{A(u)}_A(Q^2) & \propto \int \mathrm{d}x\, \Bigg[\frac{1}{9} \left\{ \left(\frac{M_N m_u}{M_{ud}} \frac{\bar{x}}{x}+\frac{M_{ud}}{\bar{x}}\right)^2  -\kappa^2 x \bar{x} \left(\frac{M_N\bar{x}}{M_{ud} x}\right)^2\left(1-\frac{q_\perp^2\bar{x}}{4\kappa^2 x}\right)\right\}  \nonumber \\
& + \frac{2}{9} \left\{-\left(M_N+\frac{m_u}{x}\right)^2+\kappa^2\frac{\bar{x}}{x} \left(1-\frac{q_\perp^2\bar{x}}{4\kappa^2 x}\right)\right\}\Bigg]  \,\exp\left(-\frac{\bar{x}q_\perp^2}{4\kappa^2 x}\right) \exp\left(-\frac{xM_{ud}^2+\bar{x}m_u^2}{\kappa^2 x\bar{x}}\right) \;,
\end{align}
and
\begin{align}
 G^{A(d)}_A(Q^2) &\propto \int \mathrm{d}x \,\Bigg[\frac{2}{9} \left\{ \left(\frac{M_N m_u}{M_{uu}} \frac{\bar{x}}{x}+\frac{M_{uu}}{\bar{x}}\right)^2  -\kappa^2 x \bar{x} \left(\frac{M_N\bar{x}}{M_{uu} x}\right)^2\left(1-\frac{q_\perp^2\bar{x}}{4\kappa^2 x}\right)\right\}  \nonumber \\
&  + \frac{4}{9} \left\{-\left(M_N+\frac{m_u}{x}\right)^2+\kappa^2\frac{\bar{x}}{x} \left(1-\frac{q_\perp^2\bar{x}}{4\kappa^2 x}\right)\right\}\Bigg]  \,\exp\left(-\frac{\bar{x}q_\perp^2}{4\kappa^2 x}\right) \exp\left(-\frac{xM_{uu}^2+\bar{x}m_u^2}{\kappa^2 x\bar{x}}\right) \;.
\end{align}

We normalize the Dirac and Pauli EM form factors using the quark counting rules and the quark anomalous magnetic moments, i.e., 
 \begin{equation}
 	F_1^u(0)=n_u\,,~~~ F_1^d(0)=n_d\,,
  \label{norm-F1}
  \end{equation} 
\begin{equation}
 F^{u}_2(0)= \kappa_u\,,~~~ F^{d}_2(0)= \kappa_d	\,,
 \end{equation}
 where $n_{u/d}$ is the number of $u/d$ quarks in the nucleon. The quark anomalous magnetic moments, $\kappa_{u/d}$, depend implicitly on the nucleon: $\kappa_{u/d}^{p/n}=2\kappa_p +\kappa_n=1.673$ and $\kappa_{d/u}^{p/n}=\kappa_p+2\kappa_n=-2.033$ are the anomalous magnetic moments of the $u$ and $d$ quarks respectively.
 Note that the normalization of the Dirac form factor allows us to fix the normalization of the wavefunctions according to the quark counting rules. We then use the same normalized wavefunctions to predict the axial form factor.
  Consequently, the normalization of the axial form factor $G^f_A(0)$, i.e. the axial charge comes out as a prediction. 

%%%%%%%% 
\begin{figure}[htbp]
\includegraphics[width=8cm]{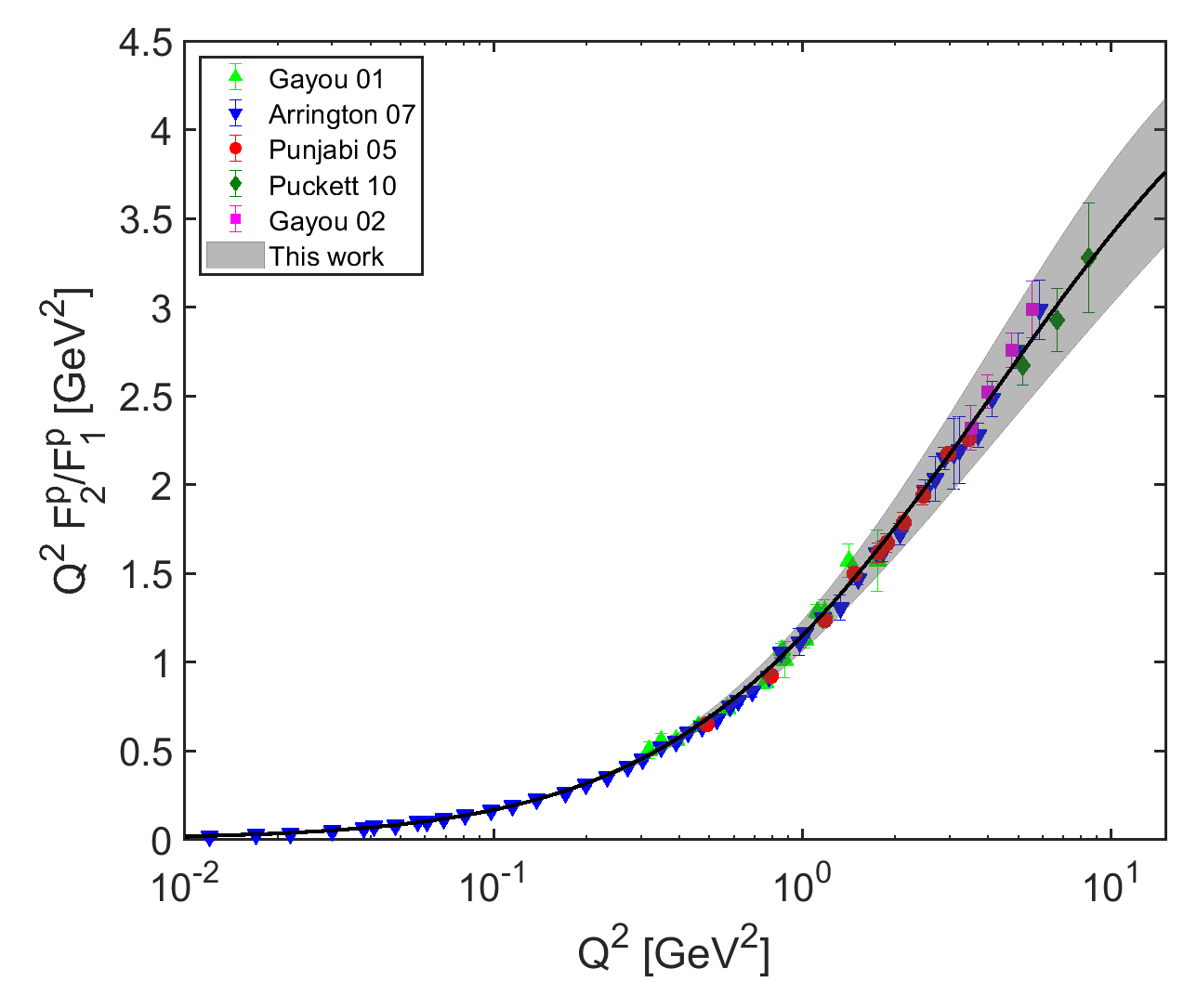}
\includegraphics[width=8cm]{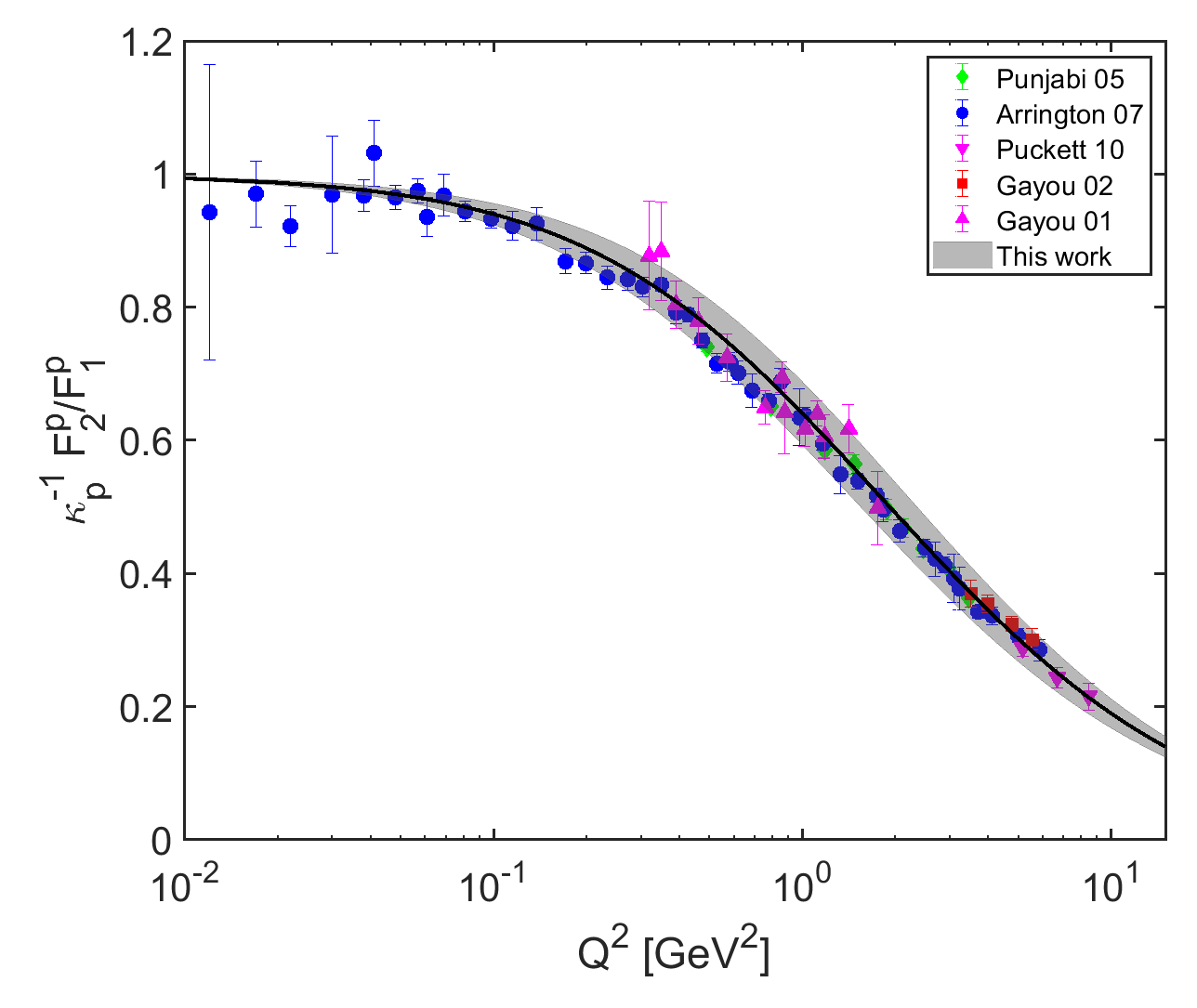}
\caption{Adjusting the quark and diquark masses to fit the data on the Pauli-to Dirac form factors ratio. The resulting masses are: $M_{uu}=0.63 \pm 0.03$ GeV and $M_{ud}=0.80 \pm 0.04$ GeV with $m_u=m_d=0.40 \pm 0.02$ GeV. The experimental data are taken from the Refs.~\cite{Gayou:2001qt,Gayou:2001qd,Arrington:2007ux,Punjabi:2005wq,Puckett:2010ac}.}
\label{fig_FFPR}
\end{figure} 

\begin{figure}[htbp]
\includegraphics[width=8cm]{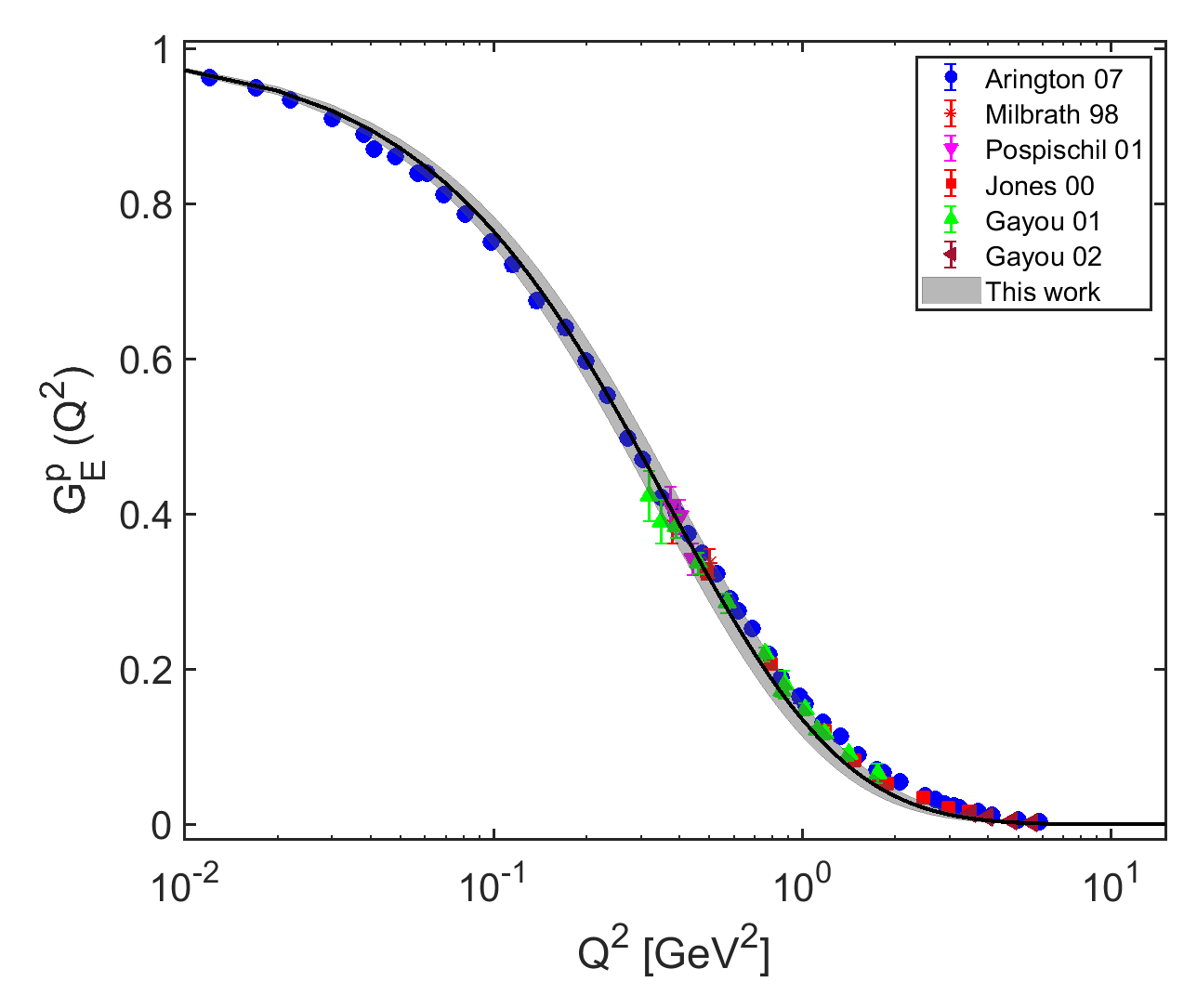}
\includegraphics[width=8cm]{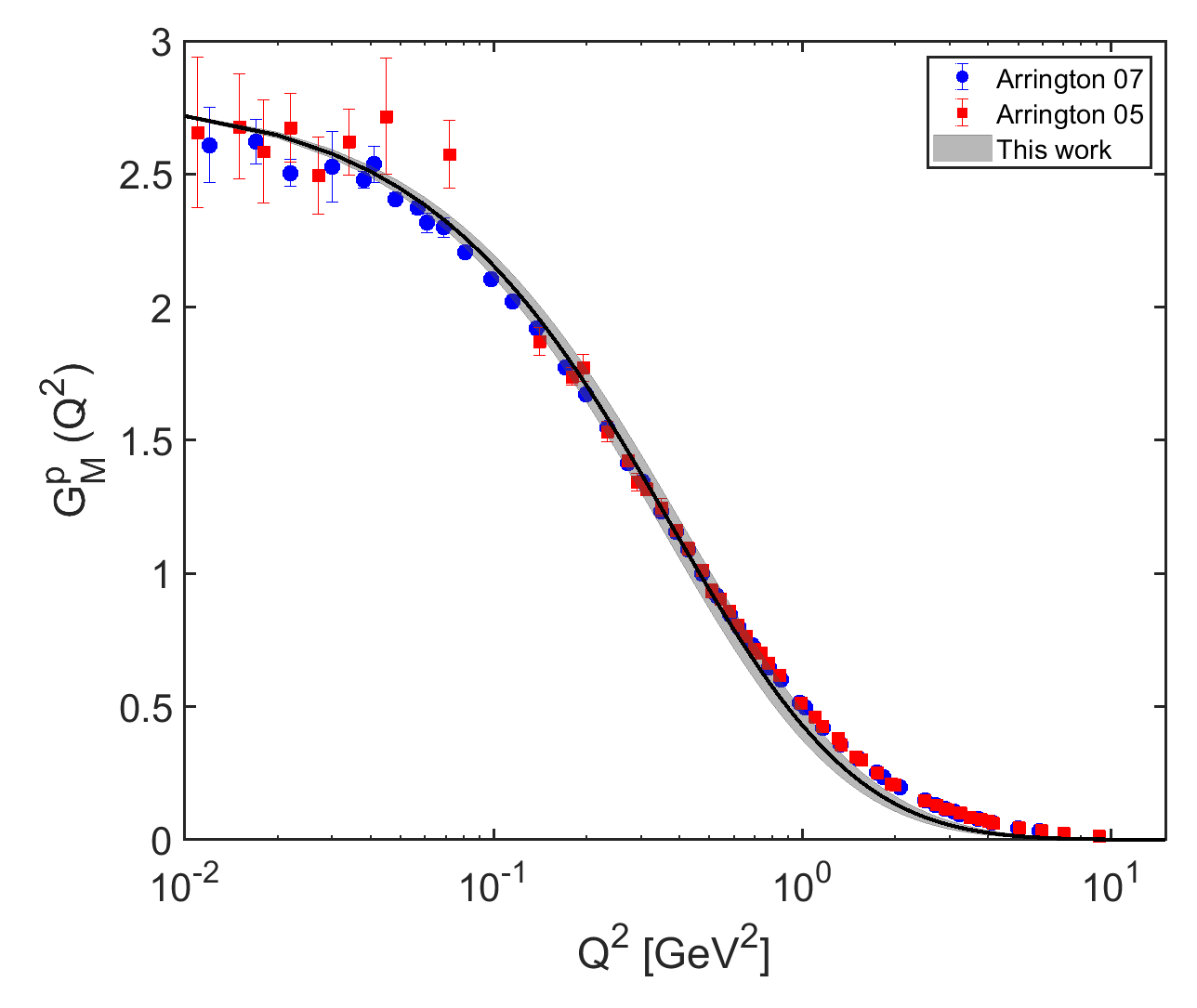}
\caption{Our predictions for the Sachs Form Factors of the proton. The grey band corresponds to the uncertainty in the holographic mass scale as well as the quark and diquark masses. The experimental data are from  Refs.~\cite{Gayou:2001qt,Gayou:2001qd,Arrington:2007ux,Milbrath:1997de,Pospischil:2001pp,Jones:1999rz} for $G_E^p(Q^2)$ and Refs.~\cite{Arrington:2007ux,Arrington:2004ae} for $G_M^p(Q^2)$.} 
\label{fig_FFP} 
\end{figure} 

\begin{figure}[htbp]
\includegraphics[width=8cm]{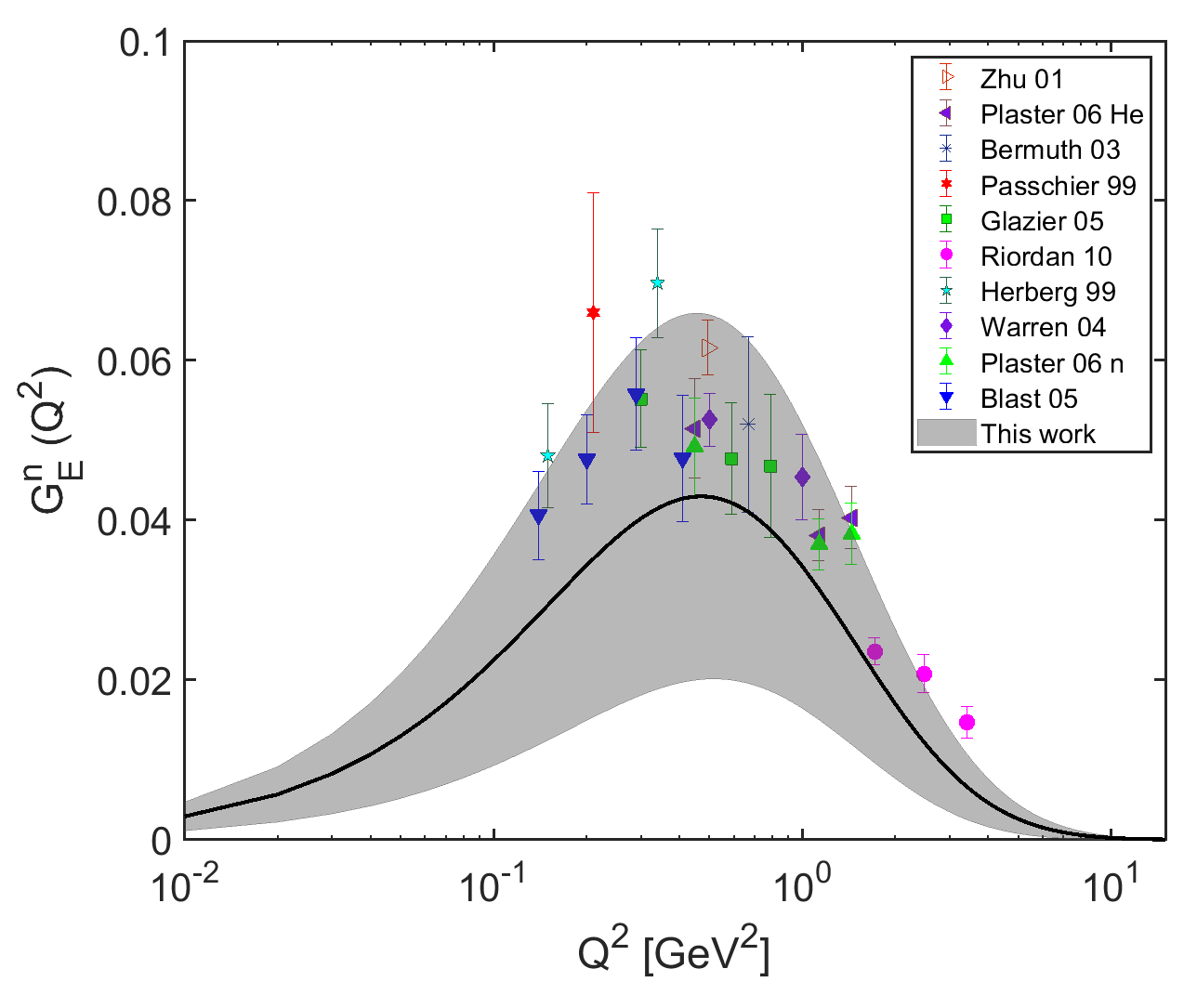}
\includegraphics[width=8cm]{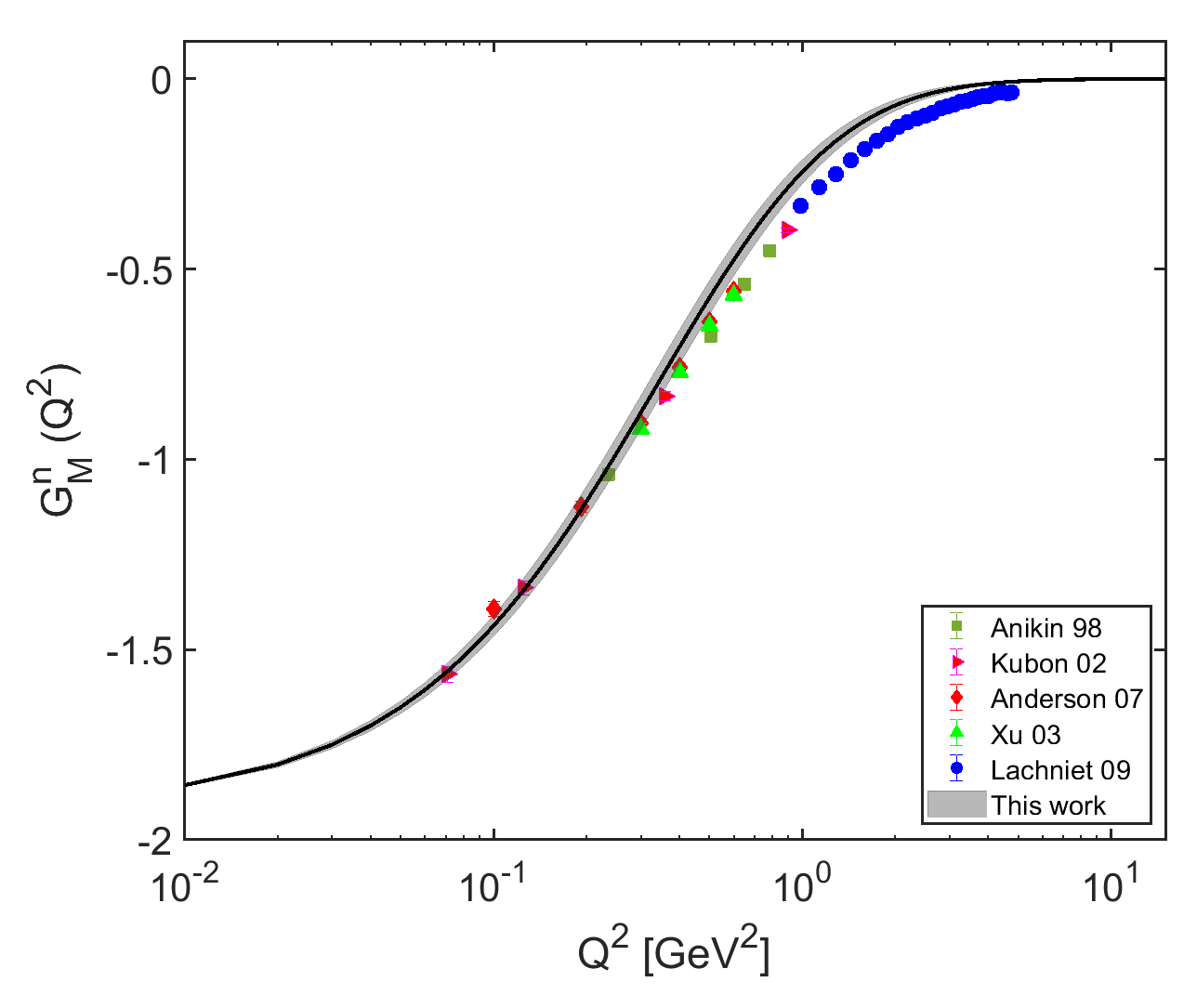}
\caption{Our predictions for the Sachs Form Factors of the neutron. The grey band corresponds to the uncertainty in the holographic mass scale as well as the quark and diquark masses. The experimental data are from   Refs.~\cite{Gayou:2001qt,Gayou:2001qd,Arrington:2007ux,Milbrath:1997de,Pospischil:2001pp,Jones:1999rz} for $G_E^n(Q^2)$, and Refs.~\cite{Anklin:1998ae,Kubon:2001rj,Xu:2002xc,Anderson:2006jp,Lachniet:2008qf} for $G_M^n(Q^2)$.} 
\label{fig_FFN}
\end{figure}

\begin{figure}[htbp]
\includegraphics[width=8cm]{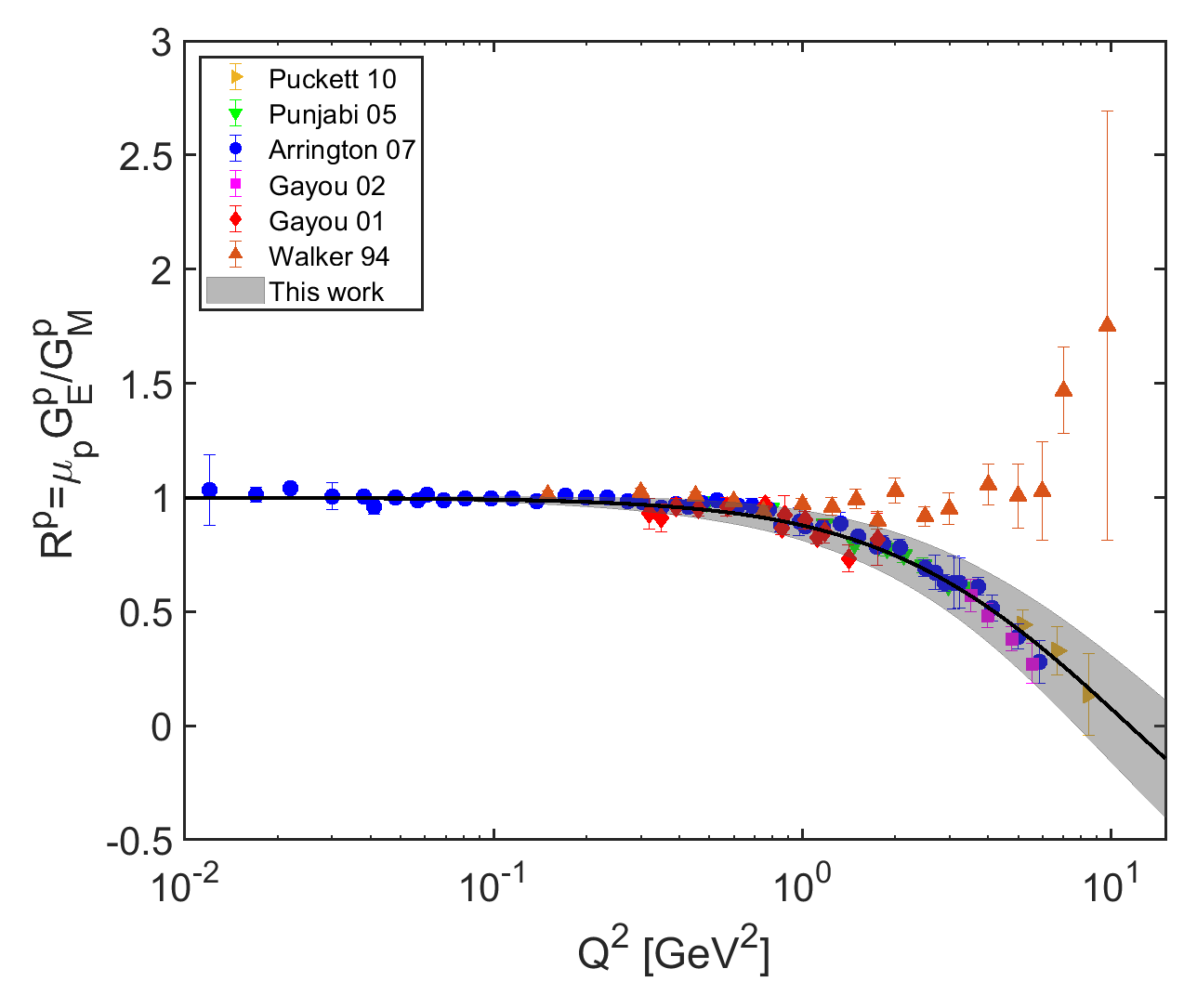}
\includegraphics[width=8cm]{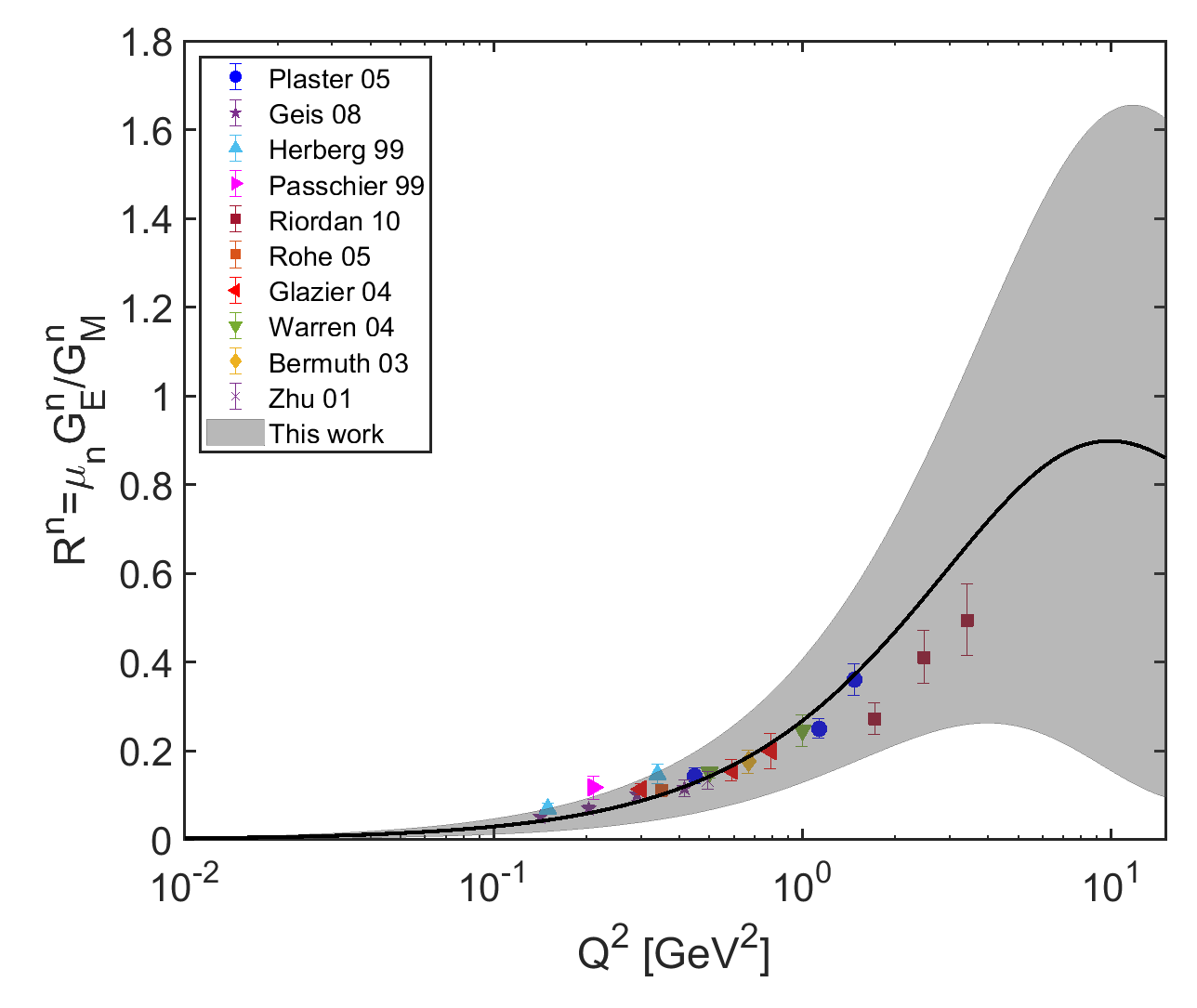}
\caption{Ratio of  Sachs form factors. Upper panel: for proton; lower panel: for neutron. The grey band corresponds to the uncertainty in the holographic mass scale as well as the quark and diquark masses. The experimental data for $R_p$ are taken from the Refs.~\cite{Gayou:2001qt,Gayou:2001qd,Arrington:2007ux,Punjabi:2005wq,Puckett:2010ac,Walker:1993vj}, while the data for $R_n$ are taken from the Refs.~\cite{Passchier:1999cj,Zhu:2001md,Warren:2003ma,Herberg:1999ud,Plaster:2005cx,Bermuth:2003qh,Geis:2007zz,Geis:2008aa,Glazier:2004ny}.}
\label{fig_FFR}
\end{figure} 

%%%%%%%% 
\begin{figure}[htbp]
\centering
\includegraphics[width=9cm,clip]{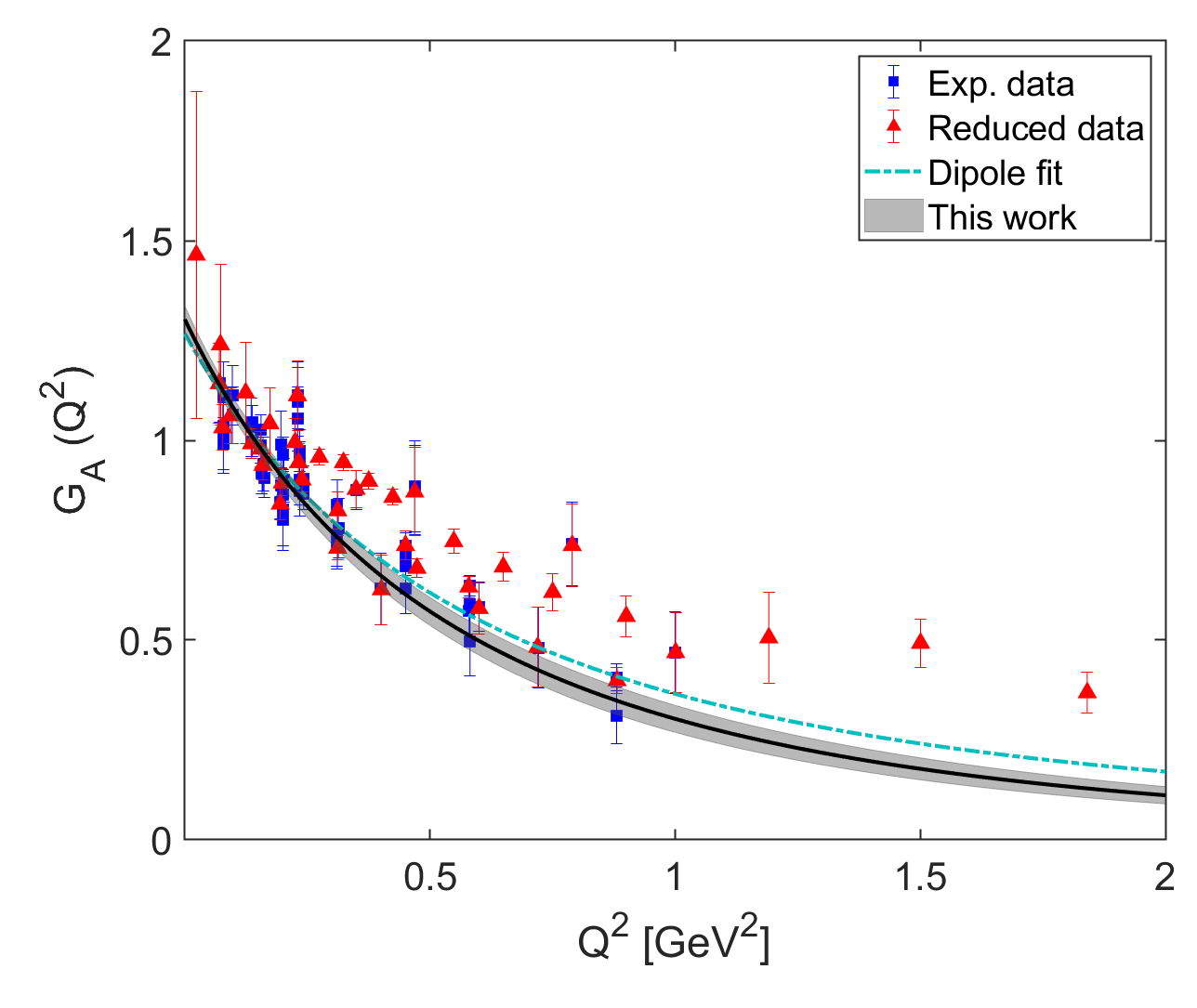}
\caption{\label{fig_AFF} Our prediction for the proton axial form factor. The grey band corresponds to the uncertainty in the holographic mass scale as well as the quark and diquark masses. The data are from  Refs.~\cite{Bernard:2001rs} (experimental data) and \cite{Hashamipour:2019pgy} (reduced data). The parametric form of the dipole fit can be found in Ref.~\cite{Bernard:2001rs}.} 
\end{figure} 
%%%%%%%%%%%

\section{Comparing to data}
To compare to experimental data, we compute the nucleon form factors using
 \be
 F_{1,2}(Q^2)= e_u F_{1,2}^u(Q^2)+e_d F_{1,2}^d(Q^2) \,,
 \label{NFF}
 \ee
 where $e_u$ and $e_d$ are the electric charge of $u$ and $d$ quarks. Note the resulting nucleon form factors satisfy $F_1(0)^p=1$, $F_1(0)^n=0$, $F_2^p(0)=\kappa_p$ and $F_2^n(0)=\kappa_n$ as expected. Similarly, the proton's axial form factor is given by
 \begin{equation}
 	G_{A}^p(Q^2)= G_{A}^u - G_{A}^d \;.
 \end{equation}
 Notice that if $M_{uu}=M_{ud}$, the axial-vector diquark contributions to the proton form factor cancel out, as found in Ref. \cite{Ma:2002ir}. Here, we consider the possibility that $M_{uu}\ne M_{ud}$. We choose to adjust the quark and diquark masses in order to fit the $F_2^p(Q^2)/F_1^p(Q^2)$ data because this ratio requires the complete set of flavour form factors. We find that $M_{uu}=0.63 \pm 0.03$ GeV and $M_{ud}=0.80 \pm0.04$ GeV with $m_u=m_d=0.40 \pm 0.02$ GeV to fit  as shown in Fig. \ref{fig_FFPR}. Recall that we use the universal AdS/QCD mass scale $\kappa=0.523 \pm 0.024$ GeV extracted from spectroscopic data \cite{Brodsky:2016rvj}. Having fixed the quark and diquark masses, all remaining predictions in this paper are generated without any further adjustment of parameters. 

We start by predicting the electric and magnetic Sachs form factors using
 \begin{equation}
 	G_E^{p/n}=F_1^{p/n}(Q^2)-\frac{Q^2}{4M_N^2} F_2^{p/n}(Q^2)\,,
 \end{equation}
 and
 \begin{equation}
 	G_M^{p/n}(Q^2)=F_1^{p/n}(Q^2) + F_2^{p/n}(Q^2) \;.
 \end{equation}
 Our predictions for the proton's Sachs form factors are shown in Fig. \ref{fig_FFP}. As can be seen, the agreement with the data is excellent in the non-perturbative, $Q^2 \le 1~\mathrm{GeV}^2$ region, where the holographic wavefunction is expected to be accurate. The predictions are less impressive for the neutron's Sachs form factors. As shown in Fig. \ref{fig_FFN}, the relatively large theory uncertainty reflects a near-exact cancellation between the active-$u$ and active $d$-quark contributions to  $G_E^n(Q^2)$. This cancellation occurring at leading order (in $\alpha_{\mathrm{em}}$) implies that the neutron form factor may be sensitive to higher order contributions which we do not consider here. We also predict the ratio $R_p=\mu_p G_E^p/G_M^p$ as shown in Fig.~\ref{fig_FFR}. Our predictions are in agreement with the polarization data~\cite{Gayou:2001qt,Gayou:2001qd,Arrington:2007ux,Punjabi:2005wq,Puckett:2010ac} but are not consistent with the data obtained using the Rosenbluth separation method~\cite{Walker:1993vj}. This is expected since our model parameters are adjusted to fit the polarization data of the $F_2^p(Q^2)/F_1^p(Q^2)$ only as shown in Fig.~\ref{fig_FFPR}.

We predict the root-mean-square charge and magnetic radii of the nucleons using
\begin{equation}
	\langle r_E^2 \rangle = \left. -6 \frac{\mathrm{d}G_E}{\mathrm{d}Q^2} \right|_{Q^2=0}; ~~ \langle r_M^2 \rangle = \left. -\frac{6}{G_M} \frac{\mathrm{d}G_M}{\mathrm{d}Q^2} \right|_{Q^2=0} \,,
\end{equation}
respectively. As shown in Table \ref{radii}, our predictions are in very good agreement with the experimentally measured values. This is consistent with the fact our predictions are most reliable in the non-perturbative, low $Q^2$ regime. Note that the most recent measurements of the proton's charge radius based on the lamb shift in atomic hydrogen \cite{Bezginov:2019mdi}, muonic hydrogen \cite{Pohl:2010zza} and the high precision electron-proton scattering experiment at JLAB  \cite{Xiong:2019umf} are all consistent with each other 
and smaller than the earlier CODATA  extraction, $\langle r_E\rangle_p=0.8768(69)$ fm \cite{Mohr:2008fa}. As shown in Table \ref{radii}, our prediction for the charge radius of the proton is in agreement with the recent measurements. In fact, the agreement with data extends also to the magnetic radius of the proton as well as the electric and magnetic radius of the neutron.

\begin{table*}[htp]
%\caption{default}
\begin{center}
\begin{tabular}{|c|c|c|}
\hline
  Radius & Our prediction & Experimental data\\
 \hline
 $\langle r_E\rangle_p$ fm & $0.833\pm 0.010$ & $0.833 \pm 0.010$ \cite{Bezginov:2019mdi}; $0.831\pm0.019$ \cite{Xiong:2019umf} ; $0.841 \pm 0.084$ \cite{Pohl:2010zza}\\
 $\langle r_M\rangle_p$ fm  &  $ 0.7985\pm 0.0313$ &  $ 0.851\pm 0.026$ \cite{Tanabashi:2018oca}\\
 \hline
 $\langle r^2_E\rangle_n$  fm$^2$ & $-0.0704\pm 0.0434 $ & $ -0.1161 \pm 0.0022 $ \cite{Tanabashi:2018oca}; $-0.110 \pm 0.008$ \cite{Atac:2021wqj} \\
 $\langle r_M\rangle_n $ fm & $ 0.8388\pm 0.0288 $ &  $0.864^{+0.009}_{-0.008} $ \cite{Tanabashi:2018oca}\\
 \hline
\end{tabular}
\end{center}
\caption {Our predictions for the electric and magnetic radii of the nucleons, compared to the experimental measurements~\cite{Bezginov:2019mdi,Xiong:2019umf,Tanabashi:2018oca}.} 
\label{radii}
\end{table*}%

We finally predict the axial form factor and axial charge for the proton. We find $g_A=G^p_A(0)=1.303\pm0.033$, which is to be compared with the experimental value ~\cite{Tanabashi:2018oca},  $g_A=1.2695\pm 0.0029$, and the lattice prediction \cite{Chang:2018uxx}, $g_A=1.271\pm 0.013$ , as well as a more recent lattice calculation with higher statistics \cite{Gupta:2018qil}, $g_A=1.218(25)(30)$.  The radius of the proton evaluated using the axial form factor is $0.675\pm 0.030$ fm,  which is in better agreement with experimental value $0.667\pm 0.120$ fm \cite{Hill:2017wgb}  than the recent lattice result $0.512\pm 0.034$ fm \cite{Yao:2017fym}. The $Q^2$-dependence of the axial form factor for the proton is shown in Fig.~\ref{fig_AFF}. As for the EM form factors, we see that agreement is very good in the low $Q^2 \le 1~\mathrm{GeV}^2$ region.

%%%%%%%%%%%
   \section{Conclusions}
%%%%%%%%%%%%
We have constructed spin-improved holographic light-front wavefunctions for the nucleons  and used them to predict their form factors and radii, with the quark and diquark masses as the only adjustable parameters. Having fixed these masses using the data on the proton's Pauli-to-Dirac form factor ratio, all our remaining predictions agree remarkably well with data at low $Q^2$, where the non-perturbative holographic wavefunction is expected to be accurate. Together with previous results \cite{Forshaw:2012im,Ahmady:2016ujw,Ahmady:2016ufq,Ahmady:2019yvo,Kaur:2020emh,Ahmady:2020mht} on light mesons, our findings suggest that the light (pseudoscalar and vector) mesons and the nucleons can be treated in a unified framework in which they share the same universal light-front holographic wavefunction.

% Acknowledgement:
\section*{Acknowledgements}
MA and RS are supported by individual Discovery Grants (SAPIN-2021-00038 and SAPIN-2020-00051) from the Natural Sciences and Engineering Research Council of Canada (NSERC). DC  is supported by Science and Engineering Research Board (India) under the Grant No. CRG/2019/000895. CM is supported by new faculty start up funding by the Institute of Modern Physics, Chinese Academy of Sciences, Grant No. E129952YR0. CM also thanks the Chinese Academy of Sciences President's International Fellowship Initiative for the support via Grants No. 2021PM0023.

%===============================
\end{document}